\documentclass[12pt,epsf]{article}
\usepackage{graphicx}
\begin{document}


\def\a{\alpha}
\def\b{\beta}
\def\c{\varepsilon}
\def\d{\delta}
\def\e{\epsilon}
\def\f{\phi}
\def\g{\gamma}
\def\h{\theta}
\def\k{\kappa}
\def\l{\lambda}
\def\m{\mu}
\def\n{\nu}
\def\p{\psi}
\def\q{\partial}
\def\r{\rho}
\def\s{\sigma}
\def\t{\tau}
\def\u{\upsilon}
\def\v{\varphi}
\def\w{\omega}
\def\x{\xi}
\def\y{\eta}
\def\z{\zeta}
\def\D{\Delta}
\def\G{\Gamma}
\def\H{\Theta}
\def\L{\Lambda}
\def\F{\Phi}
\def\P{\Psi}
\def\S{\Sigma}

\def\o{\over}
\def\beq{\begin{eqnarray}}
\def\eeq{\end{eqnarray}}
\newcommand{\gsim}{ \mathop{}_{\textstyle \sim}^{\textstyle >} }
\newcommand{\lsim}{ \mathop{}_{\textstyle \sim}^{\textstyle <} }
\newcommand{\gtrsim}{ \mathop{}_{\textstyle \sim}^{\textstyle >} }
\newcommand{\lesssim}{ \mathop{}_{\textstyle \sim}^{\textstyle <} }
\newcommand{\vev}[1]{ \left\langle {#1} \right\rangle }
\newcommand{\bra}[1]{ \langle {#1} | }
\newcommand{\ket}[1]{ | {#1} \rangle }
\newcommand{\EV}{ {\rm eV} }
\newcommand{\KEV}{ {\rm keV} }
\newcommand{\MEV}{ {\rm MeV} }
\newcommand{\GEV}{ {\rm GeV} }
\newcommand{\TEV}{ {\rm TeV} }
\def\diag{\mathop{\rm diag}\nolimits}
\def\Spin{\mathop{\rm Spin}}
\def\SO{\mathop{\rm SO}}
\def\O{\mathop{\rm O}}
\def\SU{\mathop{\rm SU}}
\def\U{\mathop{\rm U}}
\def\Sp{\mathop{\rm Sp}}
\def\SL{\mathop{\rm SL}}
\def\tr{\mathop{\rm tr}}

\def\IJMP{Int.~J.~Mod.~Phys. }
\def\MPL{Mod.~Phys.~Lett. }
\def\NP{Nucl.~Phys. }
\def\PL{Phys.~Lett. }
\def\PR{Phys.~Rev. }
\def\PRL{Phys.~Rev.~Lett. }
\def\PTP{Prog.~Theor.~Phys. }
\def\ZP{Z.~Phys. }

\newcommand{\rem}[1]{{\bf #1}}


\baselineskip 0.7cm

\begin{titlepage}

\begin{flushright}
UT-05-02\\
TU-737\\
hep-ph/0502074
\end{flushright}

\vskip 1.35cm
\begin{center}
{\large \bf
Dark Matter and Baryon Asymmetry of the Universe\\
in Large-Cutoff Supergravity 
}
\vskip 1.2cm
M. Ibe${}^{1}$, Takeo Moroi${}^{2}$ and T. Yanagida${}^{1,3}$
\vskip 0.4cm

${}^1${\it Department of Physics, University of Tokyo\\
     Tokyo 113-0033, Japan}
\vskip 0.2cm

${}^2${\it Department of Physics, Tohoku University\\
     Sendai 980-8578, Japan}
\vskip 0.2cm

${}^3${\it Research Center for the Early Universe, 
     University of Tokyo\\
     Tokyo 113-0033, Japan}

\vskip 1.5cm

\abstract{
We propose a consistent scenario of the evolution of the universe
based on the large cutoff supergravity (LCSUGRA) hypothesis of
supersymmetry breaking, where the gravitino and sfermion become as
heavy as $\sim O(1-10\ {\rm TeV})$.  With such a heavy gravitino,
baryon asymmetry of the universe can be generated by the non-thermal
leptogenesis via an inflaton decay without conflicting the serious
 gravitino problem.  We also see that, in the LCSUGRA scenario, relic
 density of the lightest superparticle becomes consistent with the WMAP
 value of the dark matter density in the parameter region required for
 the successful non-thermal leptogenesis.
 }
\end{center}
\end{titlepage}

\setcounter{page}{2}

\section{Introduction}

In a recent paper \cite{IIY} Izawa and two of us (M.I. and T.Y.) have
proposed a large-cutoff hypothesis in supergravity.  In this
hypothesis all higher dimensional operators such as quartic terms in
the K\"ahler potential at the GUT scale $M_{\rm GUT}\simeq 2\times
10^{16}$~GeV (or at the reduced Planck scale $M_G\simeq 2.4\times
10^{18}$ GeV) are suppressed by a large cutoff $M_*\sim O(4\pi M_G)$. 
Then, the sfermions and gravitino become order-of-magnitude heavier than
the gauginos and, consequently, masses of the sfermions and gravitino
are required to be significantly larger than the electroweak scale.
Even so, naturalness of the electroweak symmetry breaking can be
maintained by the focus-point mechanism \cite{focus} due to the fact
that the universality of the scalar masses at the GUT scale is
guaranteed in this scenario.  In \cite{IIY}, it was shown that the large cutoff
supergravity (LCSUGRA) scenario is well consistent with low-energy
phenomenology.  In particular, heaviness and universality of the
sfermion masses are good for suppressing dangerous supersymmetric
effects on the flavor violating processes, proton decay, and so on.
We consider that the presence of the large cutoff $M_*$ is a reflection
of a more fundamental physics beyond the GUT scale.

In this letter we study the cosmology of the LCSUGRA scenario.  In
SUGRA, there is a serious cosmological problem, that is the gravitino
problem. If the gravitino is unstable, it has a long lifetime and
decays after the big-bang nucleosynthesis (BBN) for an interesting
range of the gravitino mass, $m_{3/2}\sim 100 {\rm GeV}-10 {\rm TeV}$.
The decay products destroy light elements produced by the BBN and
hence the primordial abundance of the gravitino is constrained from
above to keep the success of the BBN. This leads to an upper bound on
the reheating temperature $T_R$ after inflation, since the abundance
of the gravitino is proportional to $T_R$.  The recent detailed
analysis derived a stringent upper bound $T_R\lsim 10^{6-7}$ GeV when
the gravitino decay has hadronic modes \cite{KKM}. This upper bound is
much lower than the temperatures required for most of thermal
baryogenesis including the leptogenesis \cite{FY}.  There have been
proposed various solutions to the above problem and, among them, we
consider that the non-thermal leptogenesis via an inflaton decay
\cite{KMY} is the most interesting and plausible.\footnote
{The Affleck-Dine (AD) baryogenesis \cite{AD} is an interesting
scenario, but this does not work in the LCSUGRA, since the K\"ahler
potential between inflatons and the AD fields is suppressed and the AD
fields acquire the Hubble-induced masses which set the AD fields at
the origin during the inflation \cite{Gaillard:1995az}.}
Importantly, even if the right-handed neutrinos are non-thermally
produced from the decay of the inflaton, success of the baryogenesis
is not automatic, as we will see.

The main conclusion of this letter is that the LCSUGRA scenario is
also advantageous in cosmology. 
In the next section, we show that a general argument on the
inflaton-decay scenario of leptogenesis gives us a lower bound on the
reheating temperature as $T_R \gsim 1\times 10^6$ GeV~\cite{Yanagida}.
Together with the constraint from the BBN \cite{KKM} we find the lower
bound on the gravitino mass $m_{3/2}\gsim 4$ TeV.  Importantly, such
heavy gravitino is a natural outcome of the LCSUGRA scenario.  In
section\ \ref{sec:darkmatter}, we show that the LCSUGRA with such a
large gravitino mass indeed explains the observed dark matter density.
In particular, the result of the present analysis shows that the
universal scalar mass of SUSY breaking, $m_0$, or equivalently the
gravitino mass $m_{3/2}$ is $4-10$ TeV which is in a special area of
so-called focus point region.  Remarkably, such a parameter region 
is also required for the success of the non-thermal leptogenesis via
the inflaton decay as mentioned above.  The last section is devoted to
discussion. 

\section{Leptogenesis via inflaton decay}
\label{sec:leptogenesis}

Let us first consider the leptogenesis in the framework of LCSUGRA.
Here, for simplicity, we consider the case where only the lightest
right-handed neutrino $N_1$ contributes to the lepton asymmetry,
assuming mass hierarchy among the right-handed neutrinos.  $N_1$ may
decay into $H_u + \ell$ or ${H}_u^* + {\ell}^*$ where $H_u$ is the
up-type Higgs and $\ell$ the lepton doublet.  These two decay channels
have different branching ratios when CP conservation is
violated. Interference between tree-level and one-loop diagrams
generates lepton-number asymmetry \cite{FY,l-asymmetry,HMY} as
\begin{eqnarray}
  \epsilon \equiv \frac{\Gamma (N_1\rightarrow H_u +\ell) - 
    \Gamma (N_1 \rightarrow {H}_u^* + {\ell}^*)} {\Gamma _{N_1}} 
  \simeq 
  -\frac {3}{8\pi}\frac{M_1}{\langle H_u\rangle ^2}
  m_{\nu_3}\delta _{\rm eff},
  \label{epsilon}
\end{eqnarray}
where $m_{\nu_3}$ is the heaviest (active) neutrino mass.  Using
$3\times 3$ neutrino Yukawa matrix $h$, the effective CP-violating
phase $\delta_{\rm eff}$ is given by
\begin{equation}
  \delta_{\rm eff} = 
  \frac{ {\rm Im}\left[h^2_{13} + \frac{m_{\nu_2}}{m_{\nu_3}}
      h^2_{12} + \frac{m_{\nu_1}}{m_{\nu_3}}h^2_{11}\right]}
       {|h_{13}|^2 + |h_{12}|^2 + |h_{11}|^2}.
\end{equation}
In deriving above expressions, the seesaw mass formula \cite{seesaw}
has been used.

With non-vanishing $\epsilon$ parameter, lepton-number asymmetry is
generated with the out-of-equilibrium decay of $N_1$.  Since
$\epsilon$ is proportional to $M_1$, $N_1$ is required to be heavy
enough to generate sufficient amount of baryon number density.  If we
consider the case where the right-handed neutrino is thermally
produced, for example, we obtain $M_1\gtrsim 10^{9-10}\ {\rm GeV}$
\cite{thermalLG}.  Since $T_R\gtrsim M_1$ in this case, we should
conclude that the thermal leptogenesis requires too high reheating
temperature to avoid the gravitino problem in the
conventional SUGRA models.

If the right-handed neutrino is non-thermally produced by the decay of
the inflaton $\Phi$ as $\Phi\rightarrow N_1+N_1$ ($\Phi\rightarrow
\tilde{N}_1+\tilde{N}_1$) \cite{KMY}, the situation changes.  In this
case, Boltzmann equations for the $B-L$ number density $n_{B-L}$ and
other quantities (i.e., the energy density of the inflaton $\rho_\Phi$
and that of radiation $\rho_{rad}$) are given by
\begin{eqnarray}
  \frac{dn_{B-L}}{dt} + 3 H n_{B-L} &=& 
  2 \epsilon \Gamma_\Phi m^{-1}_\Phi \rho_\Phi,
  \label{dot(B-L)}
  \\
  \frac{d\rho_\Phi}{dt} + 3 H \rho_\Phi &=&
  - \Gamma_\Phi \rho_\Phi,
  \\
  \frac{d\rho_{rad}}{dt} + 4 H \rho_{rad} &=&
  \Gamma_\Phi \rho_\Phi,
  \label{dot(rhorad)}
\end{eqnarray}
where $m_\Phi$ is the inflaton mass, $\Gamma_\Phi$ the decay rate of
the inflaton.  Notice that the factor 2 in the right-hand side of Eq.\
(\ref{dot(B-L)}) is due to the fact that two right-handed neutrinos
are produced by the decay of single $\Phi$.  Here and hereafter, we
adopt a mild assumption that the inflaton potential can be well
approximated by the parabolic one at the last stage of the reheating.

The inflaton decays when the expansion rate of the universe becomes
comparable to the decay rate $\Gamma_\Phi$; we define the reheating
temperature as
\begin{eqnarray}
  T_R \equiv 
  \left( \frac{10}{g_* \pi^2} M_G^2 \Gamma_\Phi^2 \right)^{1/4},
\end{eqnarray}
where $g_*$ is the effective number of the massless degrees of
freedom.  (In our numerical study, we use $g_*=228.75$.)  As well as
the entropy production, generation of the $B-L$ asymmetry is most
effective when $H\sim\Gamma_\Phi$.  Consequently, we obtain the
relation between $n_{B-L}$ and the entropy density $s$ as
\begin{eqnarray}
  \frac{n_{B-L}}{s}
  = \kappa \epsilon \frac{T_R}{m_{\Phi}}
  = 9.9 \times 10^{-11} \times
  \kappa
  \left(\frac{T_R}{10^6{\rm GeV}}\right)
  \left(\frac{2M_1}{m_\Phi}\right)
  \left(\frac{m_{\nu_3}}{0.05{\rm eV}}\right)
  \delta_{\rm eff},
  \label{n(B-L)}
\end{eqnarray}
where $\kappa$ is a constant of $O(1)$.  In the second equality, we
have taken $\langle H_u\rangle \simeq 174$ GeV.  Qualitative behavior
of Eq.\ (\ref{n(B-L)}) can be easily understood using the
instantaneous-decay approximation: $m_\phi n_{B-L}(T_R)\sim\epsilon
[\rho_\phi]_{\rm decay}\sim \epsilon g_*T_R^4$ and $s(T_R)\sim
g_*T_R^3$.  Using the order-of-magnitude relation $n_{B-L}\sim n_B$
after the spharelon transition, one can see that, in order to generate
the enough baryon asymmetry of the universe suggested by the WMAP
\cite{WMAP}
\begin{equation}
  \frac{n_B}{s} \simeq 0.9\times 10^{-10},
  \label{nB(WMAP)}
\end{equation}
$T_R$ should be higher than $\sim 10^6\ {\rm GeV}$.  

Since this lower bound is very close to the upper bound on $T_R$ from
the gravitino problem, we have performed a careful calculation of the
resultant baryon number in this scenario.  We have numerically solved
Eqs.\ (\ref{dot(B-L)}) $-$ (\ref{dot(rhorad)}) from the cosmic time
$t\ll\Gamma_\Phi^{-1}$ to $t\gg\Gamma_\Phi^{-1}$.  Then, we have
calculated the entropy density at $t\gg\Gamma_\Phi^{-1}$ using the
relations 
\begin{eqnarray}
  \rho_{rad} = \frac{\pi^2}{30} g_* T_R^4, ~~~
  s = \frac{2\pi^2}{45} g_* T_R^3.
\end{eqnarray}
Taking the ratio of $n_{B-L}$ to $s$, we have obtained the $\kappa$
parameter in Eq.~(\ref{n(B-L)}) as
\begin{eqnarray}
\kappa\simeq 2.44.
\end{eqnarray}
Using $n_B=\frac{28}{79}n_{B-L}$ \cite{turner}, baryon-to-entropy ratio
is given by 
\begin{eqnarray}
  \frac{n_{B}}{s}
  \simeq 8.2 \times 10^{-11} \times
  \left(\frac{T_R}{10^6{\rm GeV}}\right)
  \left(\frac{2M_1}{m_\Phi}\right)
  \left(\frac{m_{\nu_3}}{0.05{\rm eV}}\right)
  \delta_{\rm eff}.
\end{eqnarray}
Combining this with Eq.\ (\ref{nB(WMAP)}), we derive a constraint on the 
reheating temperature,
\begin{equation}
  T_R \gsim 1\times 10^{6} {\rm GeV},
  \label{TR(leptogen)}
\end{equation}
for the neutrino mass suggested from the atmospheric neutrino
oscillation experiments, $m_{\nu_3}\simeq \sqrt{\Delta m^2_{23}}\simeq
0.05$ eV.  Here, we used $M_1=\frac{1}{2}m_\Phi$, which is the maximal
possible value of $M_1$ so that the decay process $\Phi\rightarrow N_1
+N_1$ is kinematically allowed.  From the point of view of the
gravitino problem, such reheating temperature is quite dangerous.

It is notable that, in the LCSUGRA scenario where the gravitino is
quite heavy, the present non-thermal leptogenesis scenario becomes
viable in a wide parameter region.  Indeed, the recent detailed analysis
of the gravitino problem suggests that $T_R\sim 1\times 10^{6} {\rm
GeV}$ is consistent with the BBN constraints if the gravitino is heavier
than $4\ {\rm TeV}$ \cite{KKM}.  In the LCSUGRA scenario, such a large 
gravitino mass can be realized without conflicting the naturalness of
the electroweak symmetry breaking (the focus point mechanism).  In
addition, as we will discuss in the next section, when $m_{3/2}\gtrsim
4\ {\rm TeV}$, the predicted relic density of the LSP well agrees with
the currently observed dark matter density.

Before closing this section, some discussion on the constraints from
the gravitino problem may be relevant.  For the gravitino mass we are
interested, overproduction of D gives the most stringent upper bound
on $T_R$. Ref.\ \cite{KKM} used averaged value of the
observational values of D/H to set the bounds.  If we adopt the
largest value of observed D/H, upper bound on $T_R$ becomes larger by
factor $2-3$.  If so, the non-thermal leptogenesis may be consistent with
the gravitino mass smaller than $4\ {\rm TeV}$.  However, notice that,
in deriving Eq.\ (\ref{TR(leptogen)}), we considered the extreme case
where $M_1=\frac{1}{2}m_\Phi$ and $\delta_{\rm eff}=1$; in particular,
if $M_1$ is smaller, higher reheating temperature is required to
generate large enough baryon asymmetry.\footnote
{In addition, the primary purpose of Ref.\ \cite{KKM} was to derive
conservative constraints taking account of only the effects which are
well understood.  Thus, the hadrodissociations induced by the colored
superparticles emitted from the gravitino decay were not included
since such processes are very hard to estimate.  Once they are
included, the upper bound on $T_R$ will become severer.}
Thus, the non-thermal leptogenesis scenario is severely constrained
from the point of view of the gravitino problem and the large gravitino
mass is preferred to solve the conflict. 

\section{Dark matter in LCSUGRA}
\label{sec:darkmatter}

In the previous section, we have seen that the LCSUGRA scenario is
good for the viable scenario of the non-thermal leptogenesis.  In this
section, we discuss that the LCSUGRA scenario can provide, in the
reasonable portion of the parameter space, the dark-matter
density consistent with the WMAP result.

Before discussing the behavior of the dark-matter density, we first
summarize the free parameters in our analysis.  In the LCSUGRA
scenario, all the higher-dimensional operators are assumed to be
suppressed by the inverse powers of the cutoff scale $M_*$ much larger
than the reduced Planck scale.  
Thus, as usual, all scalar particles obtain a universal SUSY breaking
mass $m_0$ through the effectively minimal K\"ahler potential,\footnote 
{There may be non-diagonal corrections to scalar mass squared from
higher dimensional terms in K\"ahler potential which are suppressed by
the large cutoff $M_*$.}
which is equal to the gravitino mass $m_{3/2}$.
The universal gaugino mass $m_{1/2}$ is obtained through higher
dimensional operators in the superpotential which is suppressed by
$M_*$ (where we assume unification of the $SU(3)\times SU(2)\times U(1)$
gauge group).
The SUSY-breaking (scalar)$^3$ couplings, so called $A$ parameters, are
also generated through higher dimensional operators in the
superpotential and they are expected to be of order the gaugino mass.%
\footnote{If the vacuum expectation value (VEV) of the field $Z$ in the
SUSY breaking sector is large as the reduced Planck scale $M_G$, we
have the $A$ terms of the order the gravitino mass $m_{3/2}$.
However,  most of the dynamical SUSY breaking models suggest 
$\vev{Z}\ll M_G$~\cite{Izawa:1996pk,Chacko:1998si}. 
Therefore, we consider $|A|\ll m_{3/2}$.
}
In the present analysis, we simply take $A=0$, since their contributions
to the following discussions are sub-dominant as long as $|A|\ll m_0$. 
In addition, we can obtain $\mu$-parameter via the Giudice-Masiero
mechanism~\cite{Giudice:1988yz}.  Then, the $\mu$-parameter is also
suppressed by the large cutoff scale and becomes of the same order of
the gaugino mass.
Notice that the hierarchy $\mu\ll m_0$ is consistent with the
radiative electroweak symmetry breaking due to the focus point
mechanism \cite{focus}.  
Notably, in this case, the so-called $B$-parameter is not a free
parameter and it is equal to the gravitino mass (i.e., $B_0=m_0$) for
$m_0\gg A$.   
We consider that all of the above parameters are given at the GUT scale
in the present analysis.
Then, we have three parameters $m_0$, $m_{1/2}$, $\mu$ ($m_0 \gg
m_{1/2},\mu$) at the GUT scale. 

At the electroweak breaking scale, the gaugino masses $m_1,m_2$ and
$m_3$ can be determined by solving the RG equations, and are roughly
given by $m_1\simeq 0.4 m_{1/2}$, $m_2\simeq 0.8 m_{1/2}$ and $m_3\simeq 
2.8 m_{1/2}$, respectively.
Notice that the RG evolution of $\mu$ from the electroweak scale to the 
GUT scale is negligible, and hence, in the followings we take $\mu$ as the
value at the electroweak breaking scale.
We see that all the scalar mass parameters other than the one of the
up-type Higgs scalar are much larger than gaugino masses and $\mu$,
thus, the gauginos and Higgsinos are much lighter than sfermions at the
electroweak scale.
Thus, the LSP is always lightest neutralino which is stable by
$R$-parity conservation.  Such lightest neutralino becomes a good
candidate of the dark matter as we see below.

Postulating that the LSP becomes the dark matter, we calculate its relic
density.  Since heavier CP-even and CP-odd Higgs bosons as well as all
the sfermions become much heavier than the neutralino, the leading
processes in the neutralino pair annihilation are through $s$-channel
exchanges of $Z$ and CP-even Higgs bosons and also through the
$t$-channel exchanges of the charginos and neutralinos
\cite{Drees:1992am}.  In this case, $\tan\beta$ dependence of the
relic density is small since the annihilation processes do not
strongly depend on $\tan\beta$ except for the CP-even Higgs
exchanges.\footnote{
 $\tan\b \equiv v_u/v_d$ is the ratio of the two VEV's
of the neutral Higgs fields, $v_{u,d} \equiv \vev{H^0_{u,d}}$. 
}
On the other hand, since we consider the situation $|\mu| \sim
m_{1/2}$, the lightest neutralino may have a sizable fraction of the
Higgsino components~\cite{Feng:2000gh} and the relic density is
sensitive to $\mu$ and $m_{1/2}$ through the Higgsino fraction of the
lightest neutralino.

\begin{figure}
 \begin{center}
  \begin{minipage}{0.42\linewidth}
   \begin{center}
   (a) $m_{\rm top}=174$~GeV
    \includegraphics[width=.95\linewidth]{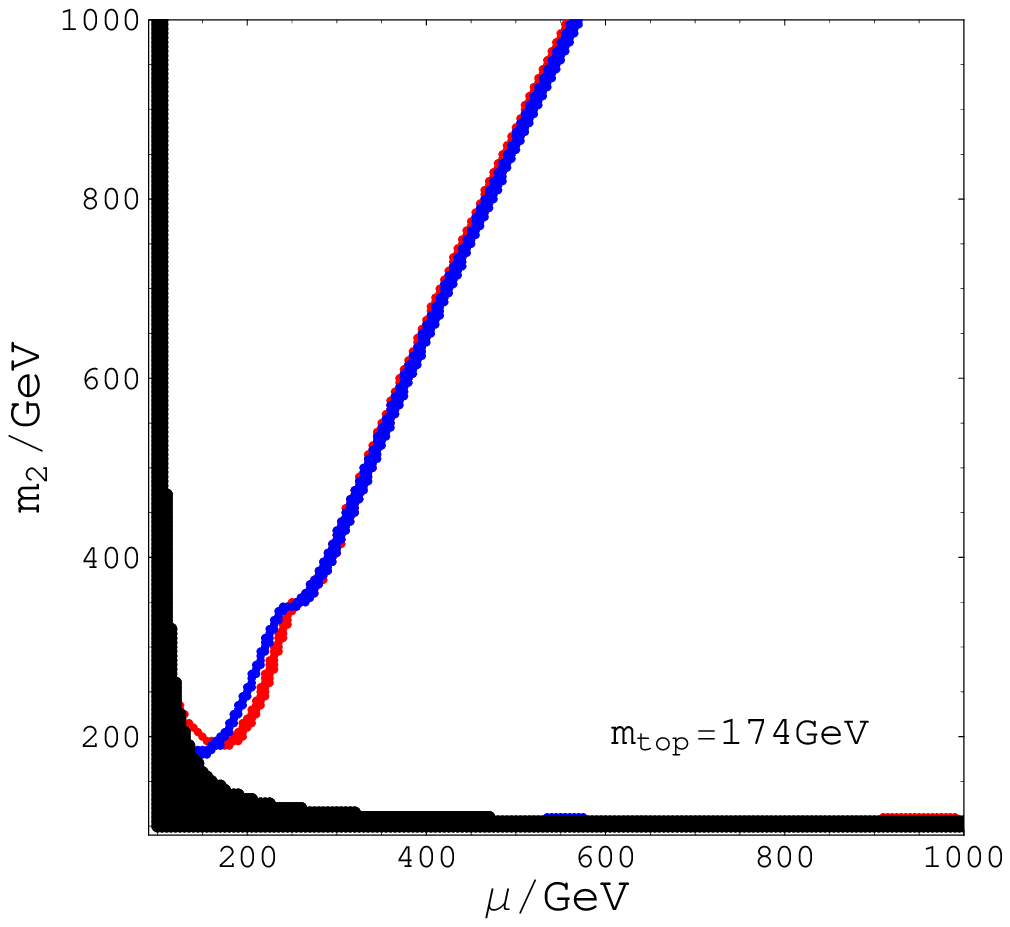}
   \end{center}
  \end{minipage}
  \begin{minipage}{0.42\linewidth}
   \begin{center}
   (b) $m_{\rm top}=178$~GeV
    \includegraphics[width=.95\linewidth]{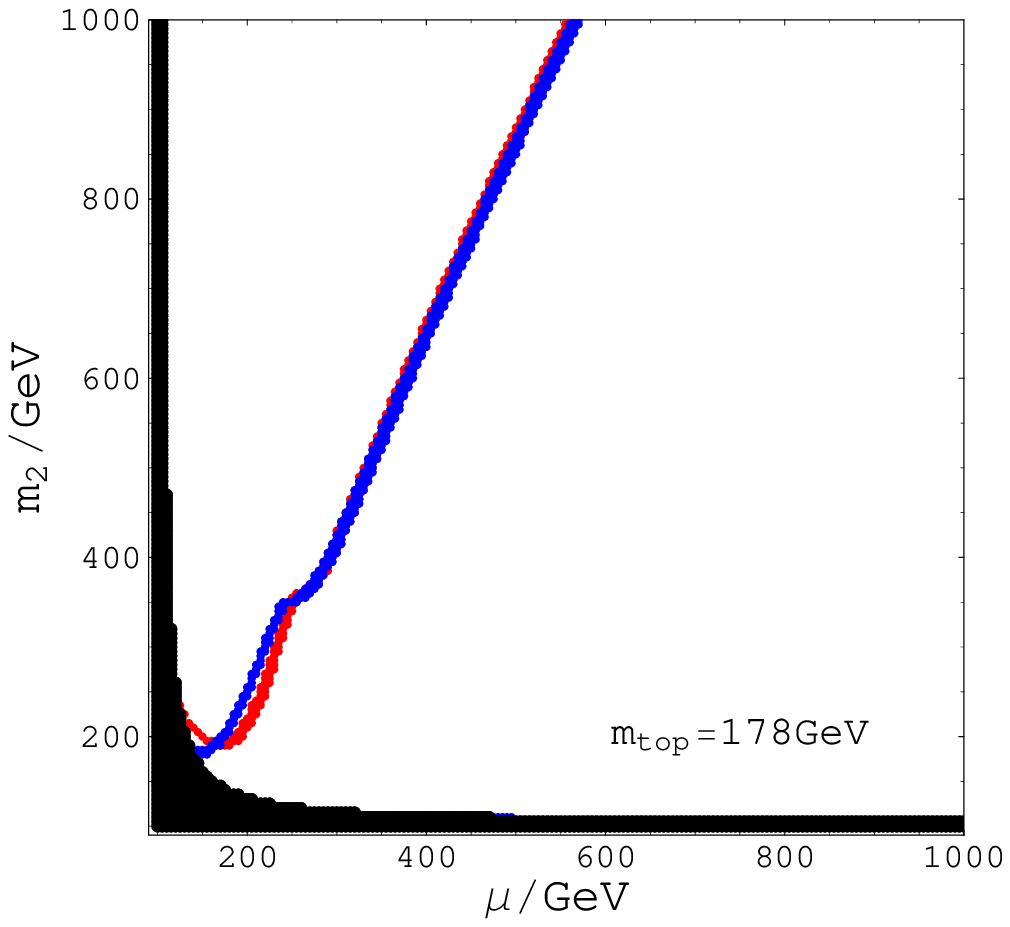}
   \end{center}
  \end{minipage}
 \end{center}
 \caption{Dark matter constraints after WMAP 
($\Omega_{\rm DM}h^2=0.1126^{+0.0161}_{-0.0181}$) 
in the ($\mu, m_2$) plane for (a) $m_{\rm top}=174$GeV and 
(b) $m_{\rm top}=178$GeV. 
The reds and blue points correspond to $\tan\beta=5$ and  $\tan\beta=30$,
 respectively. 
 The black shaded regions are excluded by the chargino mass  limit,
 $m_{\chi^{\pm}}\ge 104$GeV~\cite{chargino}. 
} 
 \label{fig:Omg}
\end{figure}

In Fig.~\ref{fig:Omg}, we show the parameter region in the ($\mu,
m_2$) plane where the neutralino density is consistent with WMAP
result~\cite{WMAP}, $\Omega_{\rm DM}h^2=0.1126^{+0.0161}_{-0.0181}$.
In this computation, we have used micrOMEGAs~1.3.0
code~\cite{Belanger:2004yn} which includes all the possible
co-annihilation effects.  In the figure, the red points are for
$\tan\beta=5$ and the blue ones for $\tan\beta=30$.  As we mentioned, 
the result is insensitive to $\tan\beta$, and we find that the relic
density is mainly determined by the parameters $\mu$ and $m_2$.  It
should be noted that the result in ($\mu, m_2$) plane is free from 
uncertainties arising from the analysis in the focus point region which
we will discuss later.  Especially, for $m_2\gsim 350$~GeV, 
$\tan\beta$ dependence of the relic density is negligible, since the
neutralino becomes heavy enough so that the annihilation process is
dominated by the channel into top-pair via $Z$-boson
exchange.\footnote
{The processes into other lighter fermion pairs are much suppressed,
since the amplitudes are proportional to the fermion masses which are
required for helicity flip in $s$-wave processes.}
The WMAP value of the relic density is realized on the line $\mu \simeq 
0.6\,m_2$, and by comparing the bino mass parameter $m_1 \simeq 0.5\,
m_2$, we find that the neutralino has a significant  Higgsino fraction
which makes annihilation more efficient.

We next reinterpret the above results by using the LCSUGRA
parameters.  In the LCSUGRA, $m_0$ and $\tan\beta$ are uniquely
determined for a given value of ($\mu, m_2$), since the SUSY-breaking
Higgs mixing parameter $B$ is determined as $B_0 = m_0$ at the GUT 
scale.

In order to understand how $\tan\beta$ depends on the LCSUGRA
parameters, it is instructive to see the tree-level condition for the
electroweak symmetry breaking although important radiative
corrections to the Higgs potential are taken into account in our
numerical calculations.  The tree-level minimization conditions of the
effective Higgs potential are given by
\begin{eqnarray}
    \frac{1}{2} m_{Z}^2
    &=& \frac{m_{H_d}^2-m_{H_u}^2 \tan^2\beta}{\tan^2\beta-1}
    - |\mu|^2, 
    \label{eq:EWSB1}
    \\
    B \mu 
    &=& \frac{\sin 2\beta}{2}(m_{H_u}^2+m_{H_d}^2+2 |\mu|^2),
    \label{eq:EWSB2}
\end{eqnarray}
where $m_{H_{u,d}}^2$ denote the soft masses squared of up-type and
down-type Higgs bosons, and all the parameters are evaluated at the
electroweak scale.\footnote
{In the actual numerical calculation, we use the running parameters at
the typical stop mass scale, where the one-loop corrections to the
Higgs potential tend to be small~\cite{deCarlos:1993yy}.  }
Since all the parameters in Eqs.\ (\ref{eq:EWSB1}) and
(\ref{eq:EWSB2}) are determined from three input parameters ($m_0$,
$m_{1/2}$, and $\mu_0$, since $B_0=m_0$), we can determine ($m_0,
\tan\beta$) for a given ($\mu, m_2$) by solving above conditions.

In order to obtain explicit relations between ($m_0, \tan\beta$) and
($\mu, m_2$), we express the parameters in Eqs.\ (\ref{eq:EWSB1}) and
(\ref{eq:EWSB2}) in terms of the input parameters as
\begin{eqnarray}
 m_{H_u}^2 &=& a_u\, m_0^2 + b_u\, |m_{1/2}|^2,\label{eq:mHu}\\
 m_{H_d}^2 &=& a_d\, m_0^2 + b_d\, |m_{1/2}|^2,\label{eq:mHd}\\
 B &\simeq & B_0=m_0,\label{eq:B}
\end{eqnarray}
where the coefficients $a_i$ and $b_i$ ($i=u,d$) are scale-dependent
functions of dimensionless gauge and Yukawa coupling constants.  Here,
$B\simeq B_0$ is the consequence of the hierarchical spectrum ($m_0\gg
m_{1/2},A$) of the LCSUGRA scenario.  As discussed in Refs.\ 
\cite{focus}, at the electroweak scale, $a_u$ is of order $10^{-2} -
10^{-1}$ while $a_d \simeq 1$ for $m_{\rm top}\simeq 170-180$~GeV,
which comes from universality of the scalar masses.  The parameter
$b_d$ becomes positive (${\cal O}(0.1)$) at the electroweak scale,
which mainly comes from positive wino and bino contributions.  On the
other hand, $b_u$ becomes negative (${\cal O}(1)$) at the electroweak
scale because of the positive gluino contributions to the stop mass
squared which affects $m_{H_u}^2$ through top Yukawa
interaction. Substituting Eqs.~(\ref{eq:mHu}), (\ref{eq:mHd}) and
(\ref{eq:B}) into the minimization conditions, we obtain explicit
relation between ($m_0,\tan\beta$) and ($\mu, m_2$) as
\begin{eqnarray}
    \mu &\simeq & \frac{a_d}{\tan\beta}\, m_0 ,
    \label{eq:mu}\\
    m_2 &\simeq &
    0.8\,m_{1/2}\simeq 0.8 \sqrt{\frac{a_u}{-(b_u+b_d)}} m_0.
    \label{eq:M2}
\end{eqnarray}
These relations show that, when $a_u \sim 10^{-2}$, the LCSUGRA
scenario is consistent with the electroweak symmetry breaking for
$\tan\beta \gsim 10$ \cite{IIY}. By using Eqs.~(\ref{eq:mu}) and
(\ref{eq:M2}), we obtain the WMAP constraint in the ($\tan\beta, m_0$)
plane from the one in the ($\mu, m_2$) plane.

\begin{figure}
 \begin{center}
  \begin{minipage}{0.42\linewidth}
   \begin{center}
   (a) $m_{\rm top} = 174$~GeV
    \includegraphics[width=.95\linewidth]{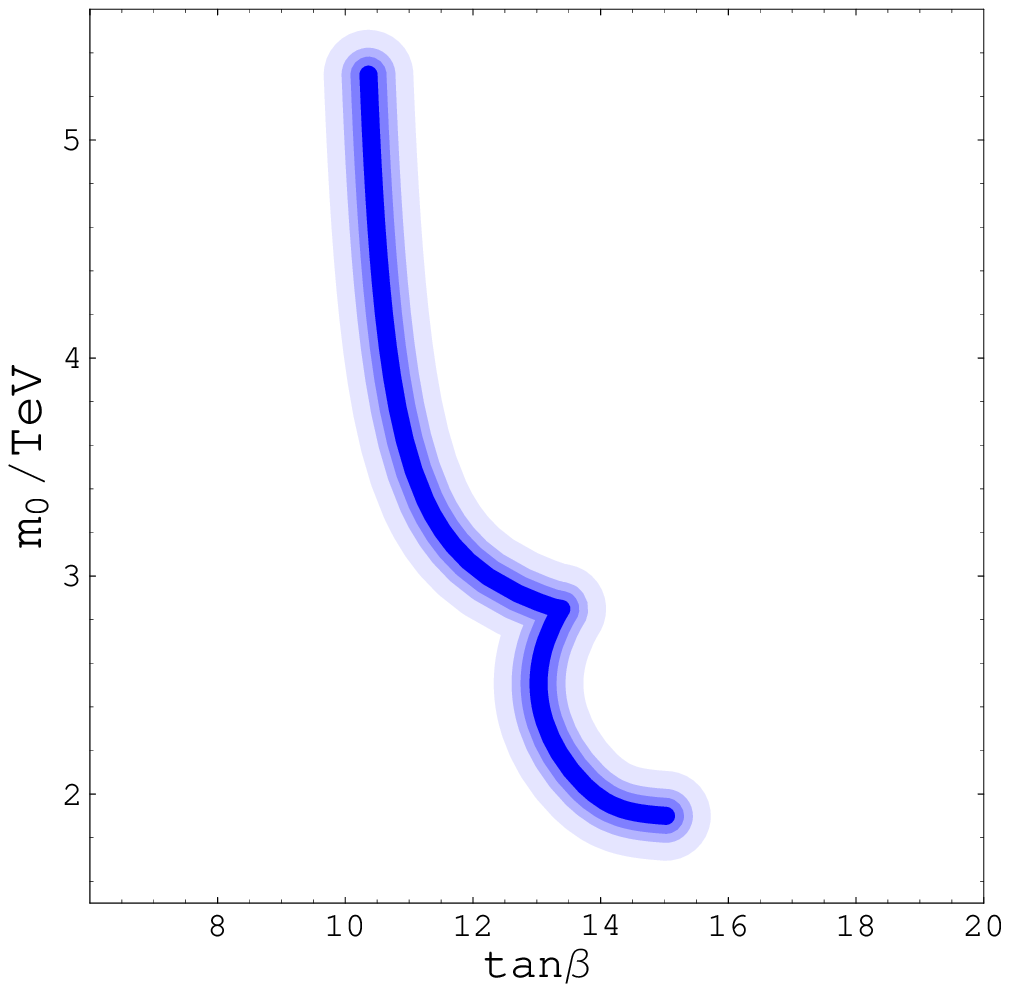}
   \end{center}
  \end{minipage}
  \begin{minipage}{0.42\linewidth}
   \begin{center}
   (b) $m_{\rm top} = 178$~GeV
    \includegraphics[width=.95\linewidth]{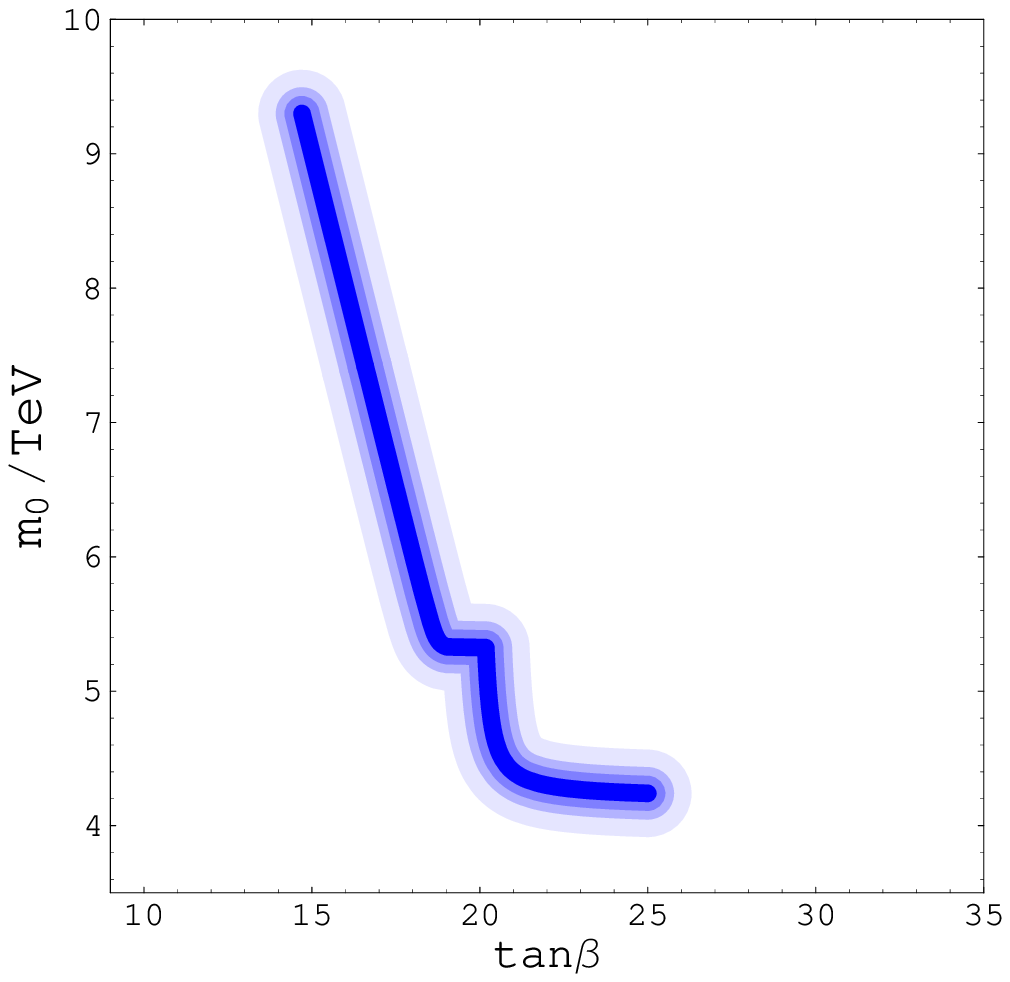}
   \end{center}
  \end{minipage}
 \end{center}
 \caption{
Cosmologically allowed regions of the relic density in the 
($\tan\beta, m_0$) plane for (a) $m_{\rm top}=174$GeV and for (b)
 $m_{\rm top}=178$GeV.~
} 
 \label{fig:tanbm0}
\end{figure}

With a more detailed numerical calculations, we relate the input
parameters to the parameters at the electroweak scale, and derive the
WMAP constraint in the ($\tan\beta, m_0$) plane.  The results are
shown in Fig.~\ref{fig:tanbm0},
 which is converted from the one in
Fig.~\ref{fig:Omg}.\footnote
{In Fig.~\ref{fig:tanbm0}, we reassure that the heavier CP-even, CP-odd 
Higgs bosons and all sfermions are much heavier than the neutralino,
which is important to obtain the results in Fig.~\ref{fig:Omg}.}
In our numerical calculations, we have used the ISAJET~7.69
code~\cite{Paige:2003mg} which takes into account the one-loop
corrections to the effective Higgs potential and the two-loop RG
evolutions of parameters.\footnote{
It should be noted that, the lines in the figures show rough fitting of
the results, since the code becomes somewhat unstable for $m_0 \gg
m_{1/2}$.     
}
In the figure, there is an upper bound on
$\tan\beta$, which is from the lower bounds on the $\mu$ and $m_2$ in
Fig.~\ref{fig:Omg}. The upper bound corresponds to parameters
$(\tan\beta,m_0,\mu,m_2) \simeq(15, 2~{\rm TeV}, 140~{\rm GeV}, 160~{\rm
GeV})$ for $m_{\rm top}=174$~GeV, and $(25, 4~{\rm TeV}, 140~{\rm GeV},
160~{\rm GeV})$ for $m_{\rm top}=178$~GeV.  
(The lower bound on the $\tan\beta$ comes from the upper bound on
$m_2\lsim 1$~TeV where we confine our attention.)  From the figure, we
find that the relic density is consistent for the WMAP result for
$m_0\gsim 2$~TeV for $m_{\rm top}=174$~GeV and $m_0\gsim 4$~TeV for
$m_{\rm top}=178$~GeV. 

As we have seen in the previous section, we find that the non-thermal
leptogenesis and the gravitino problem require $m_{3/2}\gsim 4$~TeV.
Using the fact that $m_0=m_{3/2}$ in LCSUGRA, we can see that the
LCSUGRA scenario with a relatively large gravitino mass provides a
consistent cosmological scenario which explains both baryon and dark
matter densities in the present universe.  

Finally, we comment on the uncertainties in our results which comes
from the technical difficulties in the precise calculation of the
superparticle mass spectrum in LCSUGRA.  Since some of the
superparticles become much heavier than the electroweak-symmetry
breaking scale in the focus point parameter region, an accurate
determination of the mass spectrum requires careful treatments of
various corrections.  As a result, the numerical results are somewhat
different from code to code~\cite{Allanach:2003jw,Allanach:2004jh} and
our results may be affected by such uncertainty.  To see its effect,
we have repeated our analysis with SOFTSUSY~1.9
code~\cite{Allanach:2001kg}; we have find that the changes of $m_0$ and
$\tan\beta$ for a given set of ($\mu, m_{2}$) are at most $\sim 30\ 
\%$.  Thus, we believe that our main conclusion, suggesting the LCSUGRA
scenario with a relatively large gravitino mass, does not change although 
our quantitative estimations may be slightly changed.


\section{Discussion}
\label{sec:discussion}

In this letter, we have proposed a consistent cosmological hypothesis
based on the LCSUGRA scenario of the supersymmetry breaking.  In the
LCSUGRA hypothesis, the gravitino (as well as the sfermions) becomes as
heavy as a few TeV without conflicting the naturalness of the
electroweak symmetry breaking.  If so, baryon asymmetry of the
universe can be explained by the non-thermal leptogenesis via an
inflaton decay without spoiling the success of the big-bang
nucleosynthesis.  We have also shown that, in the parameter region
suggested from the successful non-thermal leptogenesis, 
the relic density of the LSP explains naturally the dark matter density
required from the WMAP results.

\end{document}